# Non-centrosymmetric $Sr_2IrO_4$ obtained under High Pressure

Haozhe Wang[1‡], Madalynn Marshall[2‡], Zhen Wang[3], Kemp W. Plumb[4], Martha Greenblatt[2], Yimei Zhu[3], David Walker[5], Weiwei Xie[1*]

1. Department of Chemistry, Michigan State University, East Lansing, Michigan 48824, USA
2. Department of Chemistry and Chemical Biology, Rutgers University, Piscataway, New Jersey 08854, USA
3. Condensed Matter Physics and Materials Science Department, Brookhaven National Laboratory, Upton, New York 11973, USA
4. Department of Physics, Brown University, Providence, Rhode Island 02912, USA
5. Lamont Doherty Earth Observatory, Columbia University, Palisades, New York 10964, USA

‡ H.W. and M.M. contributed equally. * Email: xieweiwe@msu.edu

## *Abstract*

$Sr_2IrO_4$ with strong spin-orbit coupling (SOC) and Hubbard repulsion (U) hosts Mott insulating states. The similar crystal structure, magnetic and electronic properties, particularly the *d*-wave gap observed in $Sr_2IrO_4$ enhanced the analogies to cuprate high-$T_c$ superconductor, $La_2CuO_4$. The incomplete analogy was due to the lack of broken inversion symmetry phases observed in $Sr_2IrO_4$. Here, under high pressure and high temperature conditions, we report a non-centrosymmetric $Sr_2IrO_4$. The crystal structure and its noncentrosymmetric character were determined by single crystal X-ray diffraction and high-resolution scanning transmission electron microscopy (HR-STEM). The magnetic characterization confirms the $Ir^{4+}$ with $S = 1/2$ at low temperature in $Sr_2IrO_4$ with magnetic ordering occurred at around 86 K, where a larger moment is observed than the ambient pressure $Sr_2IrO_4$. Moreover, the resistivity measurement shows three-dimensional Mott variable-range hopping existed in the system. This non-centrosymmetric $Sr_2IrO_4$ phase appears to be a unique material to offer further understanding of high-$T_c$ superconductivity.

# Introduction

Iridates with strong spin-orbit coupling effects can generate exotic quantum phenomena, such as quantum spin liquid phases, Kitaev magnetism, and possible superconductivity.[1-4] Different from most $3d$ transition metal oxides in which the spin and orbit can be distinct in the energy scale, the spin and orbit interact heavily in $5d$ transition metal oxides. Among the Mott insulating $5d$ transition metal oxides, $Sr_2IrO_4$[5-9] has attracted significant of attention due to its similarity to cuprate high-temperature superconductor, $La_2CuO_4$[10-12]. As a single-layer Ruddlesden–Popper compound, $Sr_2IrO_4$ crystallizes in a tetragonal lattice with an inversion center ($I4_1/acd$, #142) at ambient pressure. $Sr_2IrO_4$ contains stacked $IrO_2$ square lattices where the unit cell is doubled compared to the $CuO_2$ square lattices in high-$T_c$ cuprates as a result of a staggered rotation of $IrO_6$ octahedron. Although superconductivity is not yet confirmed, many phenomena characteristic of the superconducting cuprates have been observed in electron-and hole-doped iridates including pseudogaps, Fermi arcs, and $d$-wave gaps .[13-15] The Ir-$d^5$ electrons in regular $IrO_6$ octahedron occupy the $t_{2g}$ orbitals, which can be approximated as two fully filled spin-orbital coupled $J_{eff} = 3/2$ bands and one half-filled $J_{eff} = 1/2$ band. The $J_{eff}$ band is split into an upper and lower Hubbard band by on-site Coulomb interaction. According to a previous study, as $Sr_2IrO_4$ is cooled below its Néel temperature ($T_N$, ~230 K), the spin-orbit coupled $J_{eff} = 1/2$ moments order into a basal plane commensurate Néel state. Octahedral rotations in $Sr_2IrO_4$ allow for non-zero Dzyaloshinskii-Moriya (DM) interactions that results in a canting of the ordered moments away from the crystallographic axis and a weak ferromagnetic moment per layer.[16] Such a magnetic transition maintains the inversion symmetry but lowers the rotational symmetry of the system from $C_4$ to $C_2$. However, no additional symmetry breaking has been observed by neutron or X-ray diffraction, which makes the comparison of the iridate to cuprate phenomenology incomplete. To date, multiple methods have been used to tune the Mott insulating states in $Sr_2IrO_4$, for example, isovalent Rh doping on the Ir site.[5,17-23] After partially substituting Ir with Rh, an insulator-to-metal transition can be detected. However, high pressure was also used to tune the electronic states up to 55 GPa without observing any metallic state in $Sr_2IrO_4$.[24,25]

In this report, we applied the high-pressure (6 GPa) high-temperature (1400 °C) method for synthesizing $Sr_2IrO_4$. Under such extreme conditions, the obtained $Sr_2IrO_4$ remains in a tetragonal structure but without an inversion center. The space group was determined by single-crystal X-ray diffraction (SC-XRD) as *I4mm* (#107). Unlike the ambient pressure phase, the high-pressure phase consists of the single layered $IrO_2$ square lattice, just like $CuO_2$ square in cuprate. Magnetic susceptibility measurement on high pressure $Sr_2IrO_4$ indicate a magnetic ordering temperature of approximately 86 K, which is dramatically lower than ambient pressure $Sr_2IrO_4$. Interestingly, the resistivity data shows three-dimensional Mott variable-range hopping of charge carriers between states localized by disorder with negligible long-range Coulomb interactions. Discovering the non-centrosymmetric phase in $Sr_2IrO_4$ may accelerate the realization of superconductivity and unravel the puzzle in cuprate high-$T_c$ superconductors.

## Experimental Section

**High-Pressure Synthesis.** The ambient pressure $Sr_2IrO_4$ phase was prepared accordingly by thoroughly mixing and pelletizing the materials $SrCO_3$ and $IrO_2$ and subsequently heating them to 900 °C then regrinding and reannealing at 1000 °C and subsequently reannealing at 1100 °C.[26] The ambient pressure $Sr_2IrO_4$ was pressurized to 6 GPa in 24 hours. After that, the sample was heated up to 1400 °C and stayed at 1400 °C for 4 hours. Another sample was heated to 1400 °C and stayed up to 28 hours to explore the optimal condition. The sample was cooled down to room temperature before depressurizing to the ambient pressure. The high-pressure synthesis was performed by statically compressing the sample using the Walker type multi-anvil press[27] where the original $Sr_2IrO_4$ was placed in a Pt capsule inside an $Al_2O_3$ crucible that was inserted into a Cermacast 646 octahedra pressure medium lined on the inside with a $LaCrO_3$ heater.

**Phase Analysis and Chemical Composition Determinations.** The phase identity and purity were examined using a Bruker D2 Phaser powder X-ray diffractometer with Cu K$\alpha$ radiation ($\lambda$ = 1.5406 Å). Room temperature measurements were performed with a step size of 0.004° at a scan speed of 0.55°/min over a Bragg angle (2$\theta$) range of 5–90°. FullProf Suite software[28,29] was utilized to analyze the phase information and lattice parameters from a Rietveld refinement.

**Structure Determination.** The room temperature and low temperature (100 K) crystal structure was determined using a Bruker D8 Quest Eco single crystal X-ray diffractometer, equipped with Mo radiation ($\lambda_{K\alpha}$ = 0.71073 Å) with an $\omega$ of 2.0° per scan and an exposure time of 10 s per frame. A SHELXTL package with the direct methods and full-matrix least-squares on the $F^2$ model was used to determine the crystal structure of $Sr_2IrO_4$.[30,31] To confirm the crystal structure, high-resolution scanning transmission electron microscopy (HR-STEM) images were collected and electron diffraction was conducted using a 200 kV JEOL ARM electron microscope equipped with double aberration correctors. Samples for TEM analysis were crushed in an agate mortar and deposited directly onto a holey carbon copper grid.

**Physical Properties Measurement.** Temperature and field-dependent magnetization, resistivity, and heat capacity measurements were performed with a Quantum Design physical property measurement system (PPMS) under a temperature range of 1.85–300 K and applied fields up to 9

T. Electrical resistivity measurements were accomplished with a four-probe method using platinum wires on a pelletized sample of $Sr_2IrO_4$. The polycrystalline $Sr_2IrO_4$ was pressed up to 6 GPa and heated at a lower temperature (100 °C) to eliminate the contribution of grain boundary effect but also keep the phase stable.

## Results and Discussions

**Exploring New Phase.** The new $Sr_2IrO_4$ phase (*I4mm*, #107) was formed at 6 GPa from the starting material, ambient pressure $Sr_2IrO_4$ (*I4$_1$/acd*, #142). The synthesis temperatures were set up at 1200 °C and 1400 °C. The high pressure $Sr_2IrO_4$ phase was only produced at 1400 °C. To increase the yield and grow larger crystals, the longer heating duration of 28 hours was tested. However, the secondary tetragonal phase $Sr_3Ir_2O_7$ simultaneously forms once the heating duration was increased. As a result, only 4 hours heating process can produce the specimen consisting mostly of pure phase. The resulting Le Bail fitting of the PXRD patterns for the high-pressure phase $Sr_2IrO_4$ is shown in **Fig. 1**. An overlay of the PXRD patterns in **Fig. S1** demonstrates the formation of the secondary $Sr_3Ir_2O_7$ phase. The pure phase synthesized at 1400 °C for 4 hours was used for the physical property measurements below.

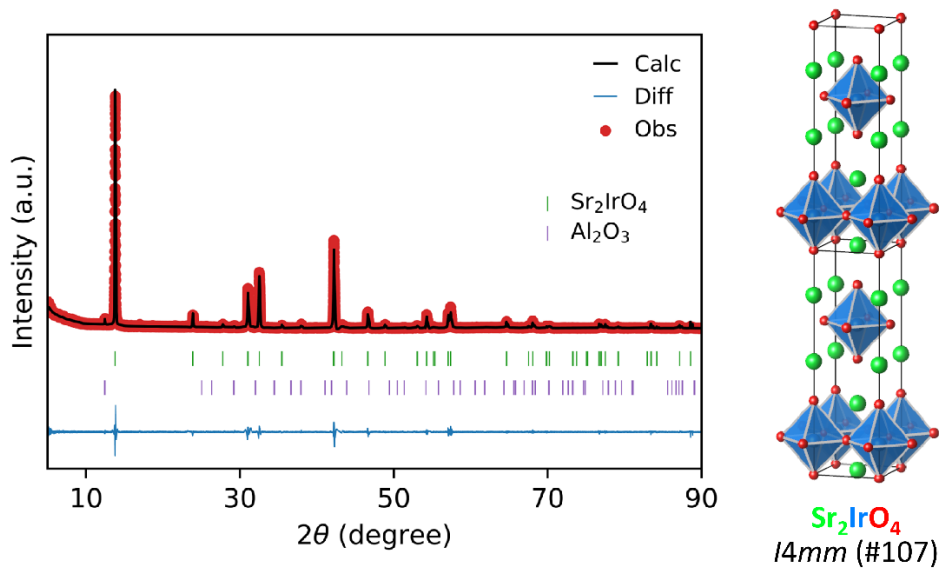

**Fig. 1 Powder X-ray diffraction pattern of the high-pressure $Sr_2IrO_4$ phase.** The experimental data (red dots) was modeled with a Rietveld refinement (black line). The blue line indicates the corresponding residual pattern (difference between observed and calculated patterns) along with Bragg peak positions for $Sr_2IrO_4$ (green) and $Al_2O_3$ (purple) represented by the vertical tick marks.

**Crystal Structure and Phase Determination.** After 4 hours of treatment at 6 GPa and 1400 °C, single crystals of Sr$_2$IrO$_4$ were formed, subsequently selected, and measured at both 300 K and 100 K using the single crystal X-ray diffractometer. High-pressure Sr$_2$IrO$_4$ crystallizes with good agreement into the tetragonal space group *I4mm*, as indicated by the single crystal X-ray diffraction (SCXRD) refinement information listed in Table S1. Similar to ambient pressure Sr$_2$IrO$_4$, the high pressure Sr$_2$IrO$_4$ phase contains the layers of IrO$_6$ octahedra with intercalated Sr atoms. The differences between these two are half-*c* lattice, the disappearance of the inversion center because of the nonsymmetric distortion of IrO$_6$ octahedra, and the disappearance of IrO$_6$ octahedral rotations in the *ab*-plane in high-pressure Sr$_2$IrO$_4$ compared to the ambient pressure phase. Shown in **Fig. 2** are crystal structures and IrO$_6$ octahedra stacking view of ambient pressure Sr$_2$IrO$_4$ (*I*4$_1$/*acd*), high-pressure Sr$_2$IrO$_4$ (*I4mm*), and previously reported La$_2$CuO$_4$ (*I4/mmm*), with Ir-O atomic distance in the IrO$_6$ octahedra highlighted. Atomic site vacancies and site disorder were considered and refined to reveal the O3 atomic site is slightly displaced from the closer ideal 4*b* site (1/2, 0, *z*) to the 8*d* site (*x*, 0, *z*) having a statistical occupancy of 0.5. The disordered model yielded a more reasonable refinement with an *R* factor of 4.35 and goodness of fit (GOF) of 1.177 while having only one O3 atomic site resulted in an *R* factor of 4.62 and GOF of 1.305. As such an angle $\delta$ can be determined from (1/2 ± $\delta$, 0, *z*) with respect to an IrO$_6$ octahedra where the O3 atoms occupy the 4*b* site. This structural disorder has been thoroughly discussed for the ambient pressure Sr$_2$IrO$_4$ structure.[32] Additionally, the high-pressure Sr$_2$IrO$_4$ phase possesses a nonsymmetric IrO$_6$ octahedra elongation along the *c* axis, ranging in Ir-O atomic distance from 1.94(6)–2.27(6) Å, as indicated in **Fig. 2b**, which is in fact the cause of noncentrosymmetric structural character. This behavior is kind of similar to the prominent feature of ambient pressure Sr$_2$IrO$_4$ that has been speculated to originate from a Jahn Teller distortion.[33-35] Previous studies under high-pressure have revealed an increase in the IrO$_6$ octahedra elongation with pressurization.[36,37] Compared to ambient pressure Sr$_2$IrO$_4$, one Ir-O along the *c*-axis is significantly elongated, with the other almost remains the same, i.e., one oxygen atom is driven away from the Ir atom, and thus the repulsion between Ir and the oxygen ligand is reduced. This will lower the energy of orbitals that contains *z* contribution and split *e*$_g$ and *t*$_{2g}$ orbitals, making the crystal field split of Ir *d* orbitals even more complicated. Together with spin-orbit coupling, this may further remove orbital degeneracies. Moreover, as pressure applied for Sr$_2$IrO$_4$, the Ir-O-Ir angle was pushed close to 180°, which is the angle in Cu-O-Cu in La$_2$CuO$_4$. The structural disorder was further confirmed at 100 K and the SCXRD refinement details can be found in SI.

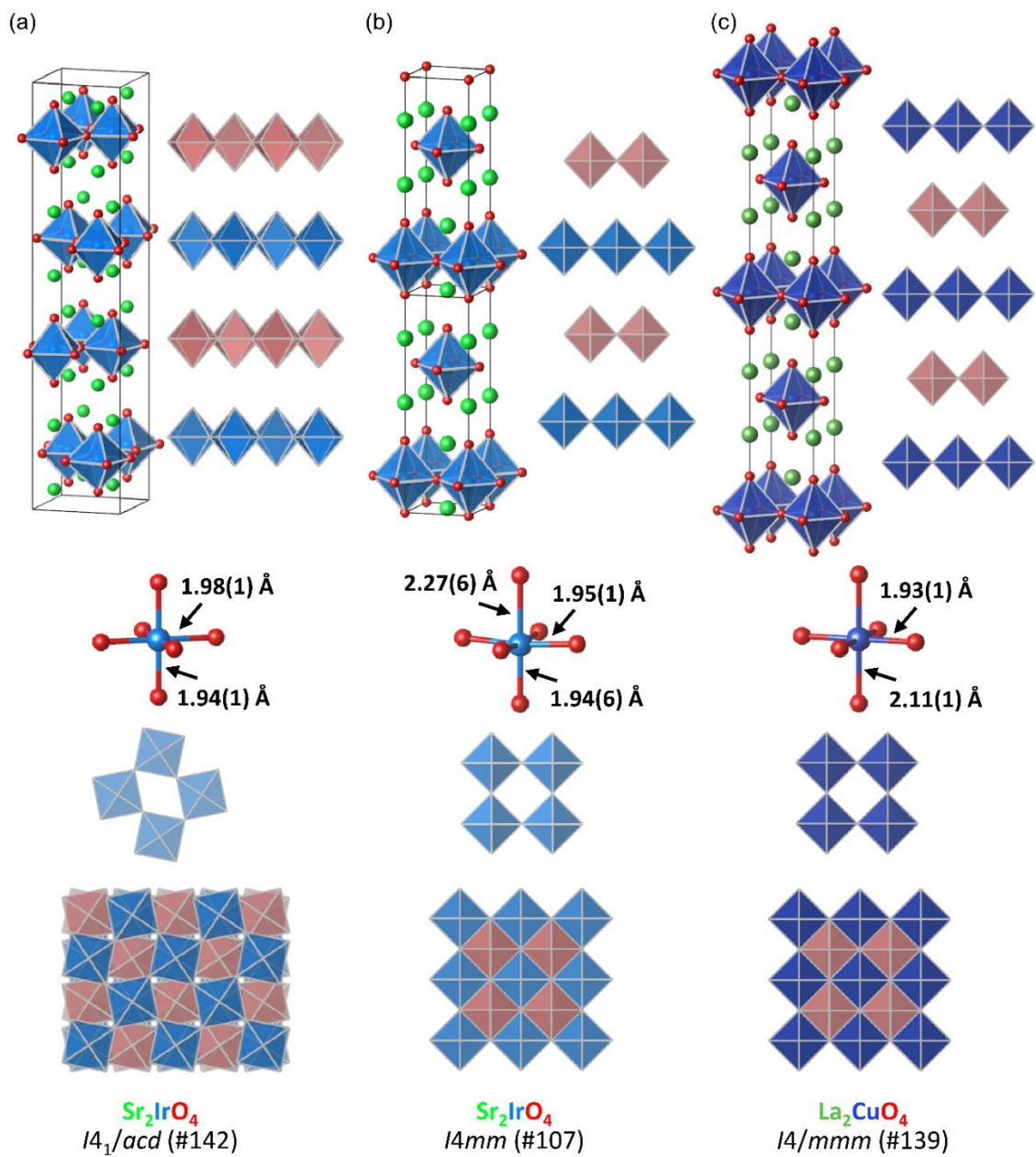

**Fig. 2 Crystal structure illustration.** Crystal structures, octahedra stacking view along *a* axis, and along *c* axis of **(a)** ambient pressure Sr$_2$IrO$_4$, **(b)** as-synthesized high pressure Sr$_2$IrO$_4$, and **(c)** previously reported La$_2$CuO$_4$, with Ir(Cu)O$_6$ octahedra and Ir(Cu)-O atomic distances presented. Green, blue, dark green, dark blue, and red atoms represent Sr, Ir, La, Cu, and O atoms, respectively. Single-layer square net is also highlighted.

**Transmission Electron Microscopy.** The non-centrosymmetric space group and loss of $IrO_6$ octahedral rotation, as well as the oxygen distortion and defects in $Sr_2IrO_4$, can at first, be surprisingly interesting, thus high-pressure $Sr_2IrO_4$ was investigated by transmission electron microscopy (TEM) to characterize its crystallographic nature. The High-angle annular dark-field scanning transmission electron microscopy (HAADF-STEM) image was obtained along the *a* axis shown in **Fig. 3a**. The TEM diffraction patterns projected down the crystalline [100] axis (**Fig. 3b**) allowed for the determination of the orientation of the images through the $d_{002}$ spacing. The *c*-axis parameter is ~12.8 Å, agreeing with the single crystal XRD results. The electron diffraction and imaging study confirmed the high quality of the nanoscale ordering in the specimen. However, the fractional spots 1/2 (110)/(1-10) were observed by TEM electron diffraction in **Fig. 3d**. As is known that $IrO_6$ tilt/rotation along the c-axis would not introduce these fractional spots. Such fractional reflection spots are related to the ordering of oxygen vacancy, which is consistent with single crystal X-ray diffraction results in **Fig. 3e**.

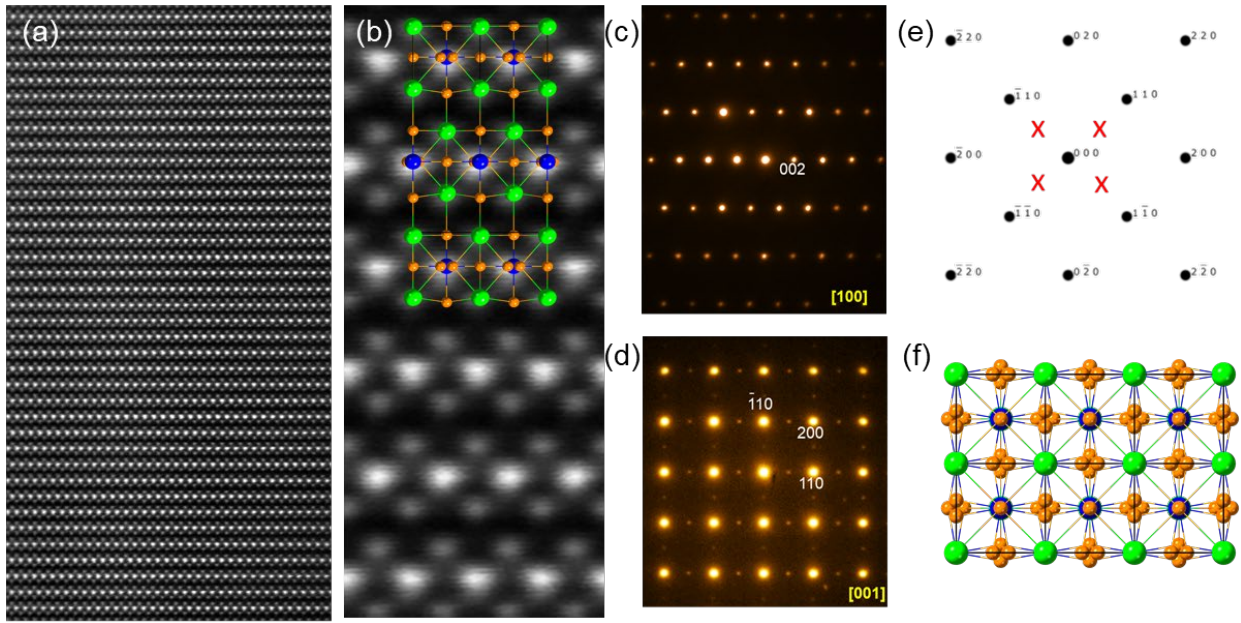

**Fig. 3 Transmission electron microscopy study of high pressure $Sr_2IrO_4$ phase. (a)** HAADF-STEM image taken along *a* axis from a large area showing the high quality of the crystal $Sr_2IrO_4$. **(b)** The zoom-in HAADF image shows the projected structure in the [100] direction, with a crystal model superimposed, where Sr (green), Ir (blue), and O (orange). **(c)** The diffraction pattern took along the [100] direction which is consistent with the simulated pattern (**Fig. 3e**) based on the crystal model determined by single crystal X-ray diffraction. **(d)** SAED pattern along the [001] direction showing fractional spots of 1/2 (110)/(1-10). **(e)** Simulated diffraction pattern and **(f)** projected crystal structure along the [001] direction based on the crystal structure determined by SCXRD. The fractional spots observed in TEM were marked in red. The single crystal structure of $Sr_2IrO_4$ with oxygen distortion was confirmed by both single crystal X-ray diffraction and TEM.

**Weak Ferromagnetic Ordering.** To study the magnetic properties of the high pressure $Sr_2IrO_4$ phase, the temperature-dependent susceptibility was measured under field cooled warming (FCW) and field cooled cooling (FCC) mode at 0.1 T shown in **Fig. 4a**. No significant differences between FCW and FCC were observed. At about 150 K, the susceptibility goes below 0, indicating a diamagnetic contribution in the system, which suggests the possible breakdown of Curie-Weiss behavior at high temperatures in the system. The data between 80–140 K was modeled with the modified Curie-Weiss law (**Eqn. 1**), shown in **Fig. 4b** and **Fig. S2b**,

$$\chi = \chi_0 + \frac{C}{T - \theta_{cw}} \tag{1}$$

where $\theta_{cw}$ is the paramagnetic Curie temperature, $\chi_0$ is the temperature independent susceptibility and $C$ is the Curie constant. From the fitting, the Curie temperature, $\theta_{cw}$, of 86(7) K was found to be comparable to the magnetic ordering temperature $T_c$ ~84 K, as determined from the minimum in the temperature derivative of $\chi$ (See **Fig. S2a** for details). The magnetic ordering temperature, consequently, decreases when compared to ambient pressure $Sr_2IrO_4$, which has a $T_c$ ~240 K.[39,40] On the other hand, it can be assumed that the Tc significantly decreases as the angle of Ir-O-Ir is more close to 180 °, which is the one observed in Cu-O-Cu in high Tc superconductor $La_2CuO_4$. The fitting also gave a negative $\chi_0$ of -2.9(9)×10$^{-3}$ emu mol$^{-1}$ Oe$^{-1}$, which provided a potential opportunity to extrapolate our Curie-Weiss fit to higher temperature. Finally, up to 160 K was included (**Fig. S2c**) and the fit yielded the effective moment $\mu_{eff}$ = 1.2(2) µB/Ir, which is more agreeable with the Hund's-rule value of 1.73 µB/Ir for $S = 1/2$ than the reported $\mu_{eff}$ = 0.33 µB/Ir for ambient pressure $Sr_2IrO_4$.

Furthermore, the magnetization of high pressure $Sr_2IrO_4$ was measured as shown in **Fig. 4c** up to 9 T at different temperatures. It appears to saturate at ~3 T at which the magnetic saturation moment ($\mu_{sat}$) was determined to be ~0.046 µB/Ir. This value is significantly lower than the theoretical value of 1/3 µB f.u$^{-1}$, however, similar to the previously reported moment for the ambient pressure $Sr_2IrO_4$ phase, which originates from spin canted antiferromagnetic (AFM) order.[39] This could also explain why the weak ferromagnetic behavior observed in the temperature dependence of magnetic susceptibility gives such a low value of moment. However, unlike the ambient pressure $Sr_2IrO_4$ phase, the magnetization reaches a maximum at around 3 T at which point the magnetization decreases. It turned out that diamagnetic transition was observed under higher fields at the respective temperatures (e.g., see the 50 K and 100 K data). At 300 K, a complete diamagnetic behavior was shown, consistent with $\chi < 0$ shown in **Fig. 4c**. Subtracting this by linearly fitting data from 7–9 T, the $\mu_{sat}$ was modified to be 0.067 $\mu_B$/Ir at 2 K and 0.014 $\mu_B$/Ir at 100 K, as presented in **Fig. 4d** and **4e**. Magnetic hysteresis was observed in the system under 2 K from -0.6 T to 0.6 T, presented in **Fig. S3**, which could be interpreted as small canting of the moments existed in the system.

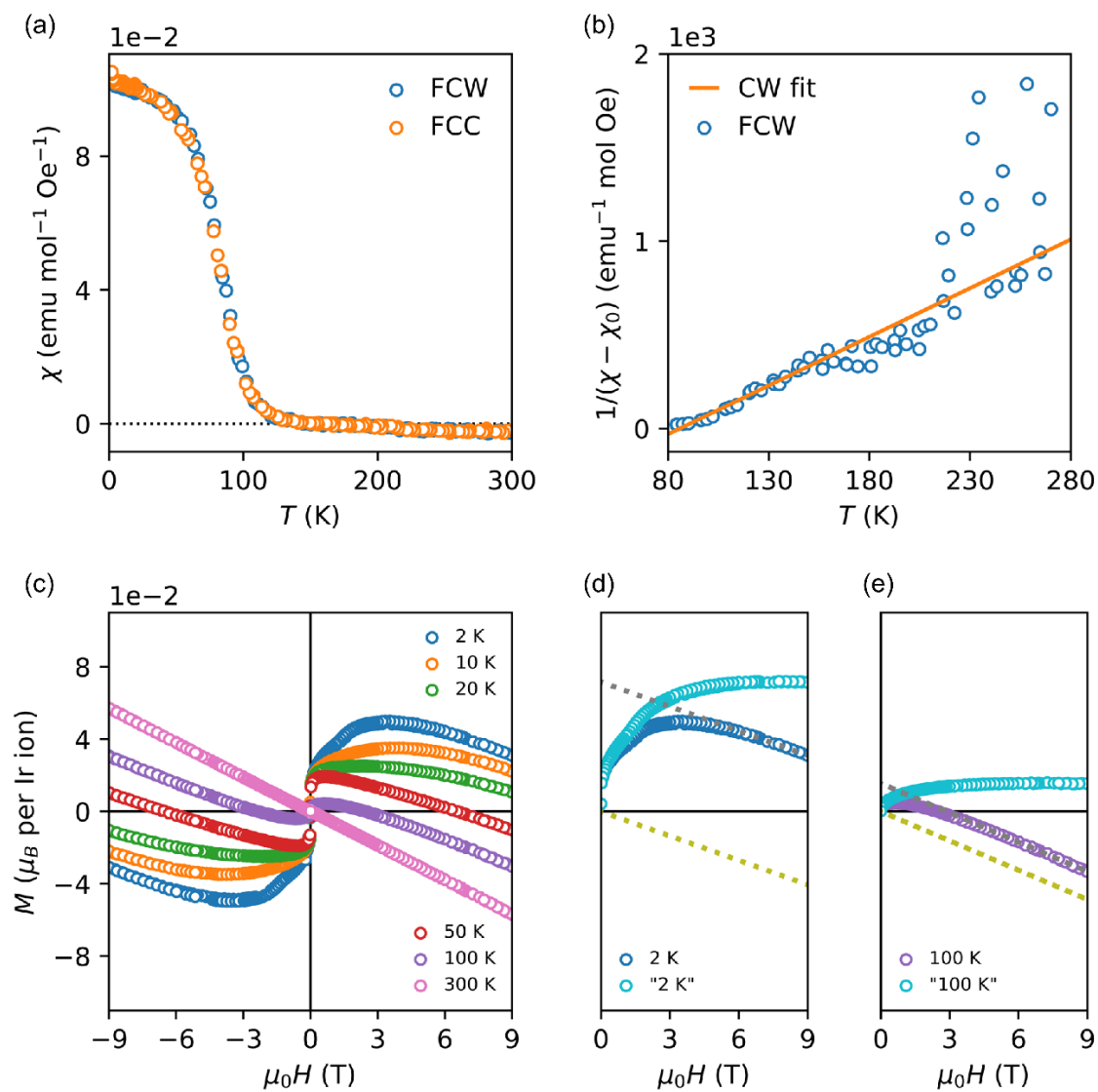

**Fig. 4 Magnetization in the dependence of temperature and field. (a)** Temperature dependence of magnetic susceptibility $\chi$ at 1000 Oe under FCW and FCC mode ranging from 2–300 K. No significant difference was observed. **(b)** The modified inverse magnetic susceptibility data (FCW, 80–140 K, blue hollow circle) fitted with the modified Curie-Weiss model (orange line). **(c)** Field dependence of magnetization up to 9 T at different temperatures. **(d)** Derivation of $\mu_{sat}$ at 2 K by linearly fitting the magnetization data from 7–9 T. **(e)** Derivation of $\mu_{sat}$ at 100 K.

**No Magnetically Induced Anomalies Observed in Specific Heat Measurement.** To confirm the magnetic transition, the specific heat over the temperature range of 2–200 K was measured under 0 T with a polycrystalline pelletized sample of $Sr_2IrO_4$, as presented in **Fig. 5*a***. Measurements under applied fields of 0.05 T and 1 T in **Fig. S3** were additionally tested to conclude no significant deviation from the 0 T specific heat. No $\lambda$ shape anomalies were observed at the whole temperature regime studied, which may result from higher temperature regions being heavily dominated by the phonon contribution. The specific heat data were fitted by the Debye model (**Eqn. 2**), and Einstein model (**Eqn. 3**), shown in **Fig. S4a** and **b**. The Debye and Einstein temperatures could then be determined as 417(2) K and 306(2) K, respectively. However, neither of these two described the experimental data well.

$$C_D = 9nR\left(\frac{T}{\theta_D}\right)^3 \int_0^{\theta_D/T} \frac{x^4 e^x}{(e^x - 1)^2} dx \qquad (2)$$

where $n$ is the number of atoms per formula unit, $R$ is the gas constant, and $\theta_D$ is the Debye temperature.

$$C_E = 3nR\left(\frac{\theta_E}{T}\right)^2 e^{\frac{\theta_E}{T}} \left(e^{\frac{\theta_E}{T}} - 1\right)^{-2} \qquad (3)$$

where $n$ is the number of atoms per formula unit, $R$ is the gas constant, and $\theta_E$ is the Einstein temperature.

The specific heat data was further fitted with two Debye model (**Eqn. 4**) and weighted Debye model (**Eqn. 5**), with and without the electronic contribution included, shown in **Fig. 5a** and **Fig. S4c**, **d**, and **e**. The data was found to be described well with two Debye model (**Fig. 5a**), and the Debye temperatures, $\theta_{D1}$ of 235(1) K, $\theta_{D2}$ of 708(5) K was obtained. At low temperatures, the first Debye mode has a larger contribution to the specific heat. Within the temperature regime studied, the expected Dulong-Petit value of $3nR$ is not recovered, and this can be explained by the high value of $\theta_{D2}$, which means that the specific heat will plateau at $T \gg \theta_{D2}$. The fitting also yields $s_{D1}$ of 3.20(3) and $s_{D2}$ of 4.51(2). The sum of these two seems a little larger than the expected value of 7 for Sr$_2$IrO$_4$, which may be attributed to the impurity of Sr$_{n+1}$Ir$_n$O$_{3n+1}$, lack of electron contribution, or overestimation of photon contribution in the model. Once the electron contribution term was included, $\theta_{D1}$ was slightly shifted to 238(2) K and the sum of $s_{D1}$ and $s_{D2}$ went down to 7.52(11).

$$C = 9s_{D1}R\left(\frac{T}{\theta_{D1}}\right)^3 \int_0^{\theta_{D1}/T} \frac{x^4 e^x}{(e^x - 1)^2} dx + 9s_{D2}R\left(\frac{T}{\theta_{D2}}\right)^3 \int_0^{\theta_{D2}/T} \frac{x^4 e^x}{(e^x - 1)^2} dx \; (+\gamma T) \qquad (4)$$

where $\theta_{D1}$ and $\theta_{D2}$ are Debye temperatures, $s_{D1}$ and $s_{D2}$ are the oscillator strengths, and $\gamma T$ is the electron contribution.

$$C = 9s_D R \left(\frac{T}{\theta_D}\right)^3 \int_0^{\theta_D/T} \frac{x^4 e^x}{(e^x - 1)^2} dx + 3s_E R \left(\frac{\theta_E}{T}\right)^2 e^{\frac{\theta_E}{T}} \left(e^{\frac{\theta_E}{T}} - 1\right)^{-2} (+\gamma T) \quad (5)$$

where $\theta_D$ and $\theta_E$ are the Debye and Einstein temperatures, $s_D$ and $s_E$ are the oscillator strengths.

It should be noted that the magnetic contribution cannot be quantitatively extracted from the specific heat data as the phonon contribution cannot be distinguished from the magnetic contribution due to the lack of a nonmagnetic analog.

At a low-temperature regime, of 2–20 K, the specific heat was measured, as shown in **Fig. S5**. The data ranging from 2–3.2 K was fitted with **Eqn. 6**, shown in **Fig. 5b**.

$$\frac{C_p}{T} = \gamma + \beta T^2 \quad (6)$$

From this fitting, a $\gamma$ and $\beta$ value of 0.0153(2) J mol$^{-1}$ K$^{-2}$ and 7.1(2) × 10$^{-4}$ J mol$^{-1}$ K$^{-3}$ corresponding to the electronic and phonon contributions to the specific heat, respectively, could be obtained. The $\beta$ value recovered the Debye temperature (**Eqn. 7**) to be 268(2) K, which is much closer to $\theta_{D1}$ rather than $\theta_{D2}$. It falls out of the temperature interval, 300–350 K, where iridates most commonly exhibit Debye temperatures.[41]

$$\theta_D = \left(\frac{12\pi^4}{5\beta} nR\right)^{\frac{1}{3}} \quad (7)$$

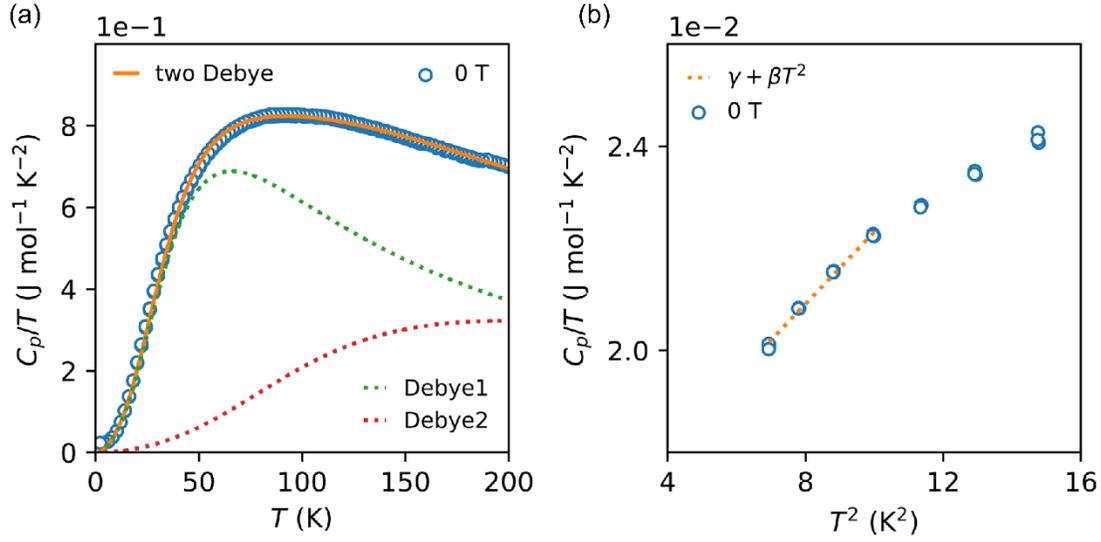

**Fig. 5 Specific heat data fitting of high pressure Sr₂IrO₄. (a)** Temperature dependence of specific heat over temperature ($C_\mathrm{p}/T$) for high-pressure Sr₂IrO₄ fitted by two Debye model in orange. Green and red dotted lines refer to the 1ˢᵗ and 2ⁿᵈ Debye model. **(b)** $C_\mathrm{p}/T$ vs $T^2$ between 2–3.2 K fitted with **Eqn. 6** (orange dotted line).

**Mott Variable-range Hopping (VRH).** It is critical to investigate the electrical conductivity in the high pressure Sr₂IrO₄ phase to compare to the Mott insulator ambient pressure Sr₂IrO₄. Temperature-dependent resistivity measurements were performed from 2–300 K with an applied field up to 9 T on a pelletized polycrystalline sample of the high pressure Sr₂IrO₄ phase, shown in **Fig. 6a**. No significant field dependence was observed, which indicates the insignificance of magnetoresistance for the high pressure Sr₂IrO₄ phase. This may be not unexpected considering the small saturation moment under fields (see the discussion above). At room temperature and 0 T, the resistivity is relatively low, only around 4 Ω cm. However, the resistivity is increases by 6 orders of magnitude upon cooling, indicating the semiconducting character of the high-pressure Sr₂IrO₄ phase.

To further analyze its behavior, we first tried to model the temperature dependence of $\rho$ with the Arrhenius law (**Eqn. 8**),

$$\rho = \rho_0 e^{E_\rho/kT} \tag{8}$$

where $\rho_0$ is the residual resistivity, $E_\rho$ is the activation energy, and $k$ is the Boltzmann constant. However, $\rho$ could not be fitted well to a $E_\rho$, shown in **Fig. S7a**, i.e., the Arrhenius law is not well obeyed. Then its temperature dependence was fitted by law in the form (**Eqn. 9**) with $v$ of 1/2 and 1/4,

$$\rho = \rho_0 e^{(T_0/T)^v} \tag{9}$$

where $\rho_0$ is the residual resistivity, and $T_0$ is the characteristic temperature. The fitting results were presented in **Fig. 6b**, and **Fig. S8**, with parameters summarized in **Table S4**. The value $v$ of 1/4 is favored over 1/2. While both of them indicate three-dimensional Mott variable-range hopping of charge carriers between localized states, the weaker temperature dependence with $v$ of 1/4 implies negligible long-range Coulomb interactions between localized electrons in the temperature regime studied. This behavior is also reported in the ambient pressure $Sr_2IrO_4$.[42] To explore the harboring quantum states in the high-pressure $Sr_2IrO_4$ phase, further examination of its transport properties is warranted.

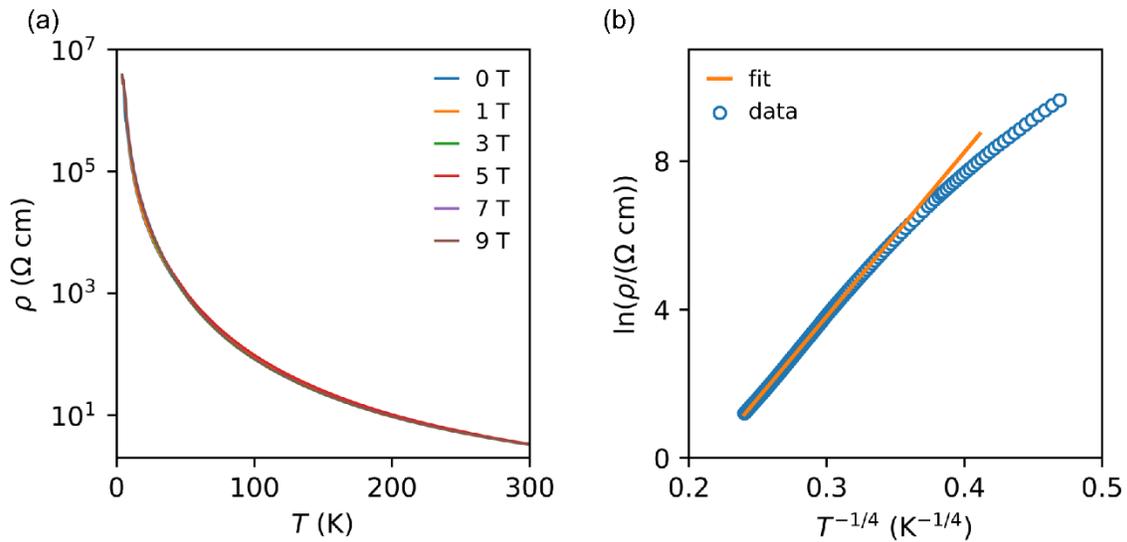

**Fig. 6 Details of field and temperature dependent resistivity. (a)** Temperature dependence of resistivity data for high-pressure phase $Sr_2IrO_4$ under fields up to 9 T. No significant derivation was observed. **(b)** The resistivity $\rho$ (blue hollow circle) ranging from 80–300 K was fitted by **Eqn. 9** with $v$ of 1/4 (orange line). A linear relationship was obtained.

## Conclusion

In summary, we reported the non-centrosymmetric $Sr_2IrO_4$ phase obtained under high pressure and high temperature conditions. The ferromagnetic ordering temperature decreases significantly to $T_c$ ~86 K from ~240 K in the ambient pressure $Sr_2IrO_4$, while there may be a possible breakdown of the Curie-Weiss law under higher temperatures. Diamagnetism was observed under room temperature and higher fields. No anomalies indicating magnetic ordering were observed in the specific heat measurements, where a greater photon contribution was obtained from the low-temperature regime. Temperature-dependent resistivity revealed three-dimensional Mott variable-range hopping of charge carriers between states localized by disorder with negligible long-range Coulomb repulsions. Further transport measurements, together with first-principal calculation, are expected to explore the electronic properties of the high-pressure $Sr_2IrO_4$ phase. Such a system may offer a promising platform to unravel the mystery of high-$T_c$ superconductivity in cuprates.

## Acknowledgments

The work at Rutgers was supported by U.S. DOE-BES under Contract DE-SC0022156. The electron microscopy work at BNL was supported by U.S. DOE-BES, Materials Sciences and Engineering Division under Contract No. DESC0012704.

## Supporting Information

Single crystal X-ray diffraction data at room temperature and 100 K; Anisotropic displacement parameters; Atomic coordinates and equivalent isotropic displacement parameters; PXRD overlay of $Sr_2IrO_4$; Magnetic susceptibility and Curie-Weiss fitting; Magnetic hysteresis; Field dependence of specific heat; Specific heat data fitted by Debye and Einstein model; Low temperature specific heat data (2–20 K); Temperature dependence of resistivity; Resistivity data fitted by Eqn. 9 with $v$ of 1/2 and 1/4; Summary of fitting parameters for resistivity data.

# Non-centrosymmetric $Sr_2IrO_4$ obtained under high pressure

Haozhe Wang[1‡], Madalynn Marshall[2‡], Zhen Wang[3], Kemp W. Plumb[4], Martha Greenblatt[2], Yimei Zhu[3], David Walker[5], Weiwei Xie[1*]

1. Department of Chemistry, Michigan State University, East Lansing, Michigan 48824, USA
2. Department of Chemistry and Chemical Biology, Rutgers University, Piscataway, New Jersey 08854, USA
3. Condensed Matter Physics and Materials Science Department, Brookhaven National Laboratory, Upton, New York 11973, USA
4. Department of Physics, Brown University, Providence, Rhode Island 02912, USA
5. Lamont Doherty Earth Observatory, Columbia University, Palisades, New York 10964, USA

‡ H.W. and M.M. contributed equally. * Email: xieweiwe@msu.edu

# Supporting Information

**Table S1** Single crystal X-ray diffraction data at room temperature and 100 K.

| Temperature | Room Temperature | 100 K |
| --- | --- | --- |
| Refined formula | $Sr_2IrO_4$ | $Sr_2IrO_4$ |
| FW (g/mol) | 431.44 | 431.44 |
| Space group | *I4mm* | *I4mm* |
| $a$ (Å) | 3.8860(5) | 3.8777(5) |
| $c$ (Å) | 12.826(2) | 12.825(2) |
| $V$ (Å$^3$) | 193.69(6) | 192.85(6) |
| Extinction Coefficient | N/A | N/A |
| $\theta$ range (°) | 3.177–33.030 | 3.177–33.075 |
| # of reflections; $R_{int}$ | 1088; 0.0627 | 1286; 0.0591 |
| # of independent reflections | 267 | 264 |
| # of parameters | 23 | 23 |
| $R_1$; $\omega R_2$ ($I > 2\delta(I)$) | 0.0409; 0.0651 | 0.0312; 0.0443 |
| Goodness of fit (GOF) | 1.177 | 1.125 |
| Diffraction peak and hole (e$^-$/ Å$^3$) | 3.658, -3.492 | 2.359, -1.96 |

**Table S2 Anisotropic displacement parameters for $Sr_2IrO_4$ at room temperature and 100 K.**

**$Sr_2IrO_4$ at Room Temperature**

| Atom | $U^{11}$ | $U^{22}$ | $U^{33}$ | $U^{23}$ | $U^{13}$ | $U^{12}$ |
|---|---|---|---|---|---|---|
| **Ir1** | -0.0018(6) | -0.0018(6) | -0.0021(6) | 0 | 0 | 0 |
| **Sr1** | 0.026(7) | 0.026(7) | 0.007(7) | 0 | 0 | 0 |
| **Sr2** | -0.001(4) | -0.001(4) | 0.005(6) | 0 | 0 | 0 |
| **O1** | 0.03(2) | 0.03(2) | -0.02(2) | 0 | 0 | 0 |
| **O2** | -0.006(10) | -0.006(10) | -0.023(18) | 0 | 0 | 0 |
| **O3** | 0.04(3) | 0.003(11) | -0.01(3) | 0 | 0.02(3) | 0 |

**$Sr_2IrO_4$ at 100 K**

| Atom | $U^{11}$ | $U^{22}$ | $U^{33}$ | $U^{23}$ | $U^{13}$ | $U^{12}$ |
|---|---|---|---|---|---|---|
| **Ir1** | -0.0004(3) | -0.0004(3) | 0.0012(7) | 0 | 0 | 0 |
| **Sr1** | 0.005(8) | 0.005(8) | 0.005(4) | 0 | 0 | 0 |
| **Sr2** | 0.002(8) | 0.002(8) | 0.000(4) | 0 | 0 | 0 |
| **O1** | 0.005(8) | 0.005(8) | -0.033(15) | 0 | 0 | 0 |
| **O2** | 0.012(10) | 0.012(10) | -0.033(14) | 0 | 0 | 0 |
| **O3** | 0.009(10) | 0.005(7) | 0.006(8) | 0 | 0.01(3) | 0 |

**Table S3 Atomic coordinates and equivalent isotropic displacement parameters for $Sr_2IrO_4$ at room temperature and 100 K.** ($U_{eq}$ is defined as one-third of the trace of the orthogonalized $U_{ij}$ tensor ($Å^2$)).

### $Sr_2IrO_4$ at Room Temperature

| Atom | Wyck. | x | y | z | Occ. | $U_{eq}$ |
|---|---|---|---|---|---|---|
| **Ir1** | 2a | 0 | 0 | 0.1513(13) | 1 | -0.0019(4) |
| **Sr2** | 2a | 0 | 0 | 0.5044(4) | 1 | 0.020(4) |
| **Sr1** | 2a | 0 | 0 | 0.79985(2) | 1 | 0.001(3) |
| **O1** | 2a | 0 | 0 | 0.328(4) | 1 | 0.013(18) |
| **O2** | 2a | 0 | 0 | 0.000(4) | 1 | -0.011(7) |
| **O3** | 8d | 0.419(9) | 0 | 0.661(7) | 0.5 | 0.010(15) |

### $Sr_2IrO_4$ at 100 K

| Atom | Wyck. | x | y | z | Occ. | $U_{eq}$ |
|---|---|---|---|---|---|---|
| **Ir1** | 2a | 0 | 0 | 0.1489(7) | 1 | 0.0001(3) |
| **Sr2** | 2a | 0 | 0 | 0.5019(4) | 1 | 0.005(5) |
| **Sr1** | 2a | 0 | 0 | 0.7978(2) | 1 | 0.002(5) |
| **O1** | 2a | 0 | 0 | 0.321(3) | 1 | -0.008(6) |
| **O2** | 2a | 0 | 0 | 0.000(3) | 1 | -0.003(8) |
| **O3** | 8d | 0.412(4) | 0 | 0.649(6) | 0.5 | 0.007(4) |

**Figure S1 Powder X-ray diffraction pattern overlay.** The experimental data of high pressure $Sr_2IrO_4$ phase synthesized at 1400 °C for ~4 hrs (black line) and ~28 hrs (red line) were presented. Bragg peak positions are indicated as $Sr_2IrO_4$ and $Sr_3Ir_2O_7$ with green and purple vertical tick marks, respectively.

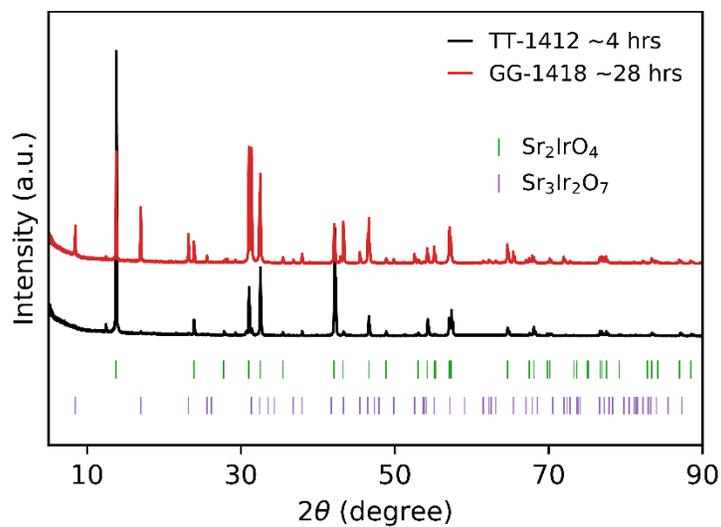

**Figure S2 Magnetic susceptibility and Curie-Weiss fitting.** (a) Temperature derivative of magnetic susceptibility $\chi$ at 1000 Oe under FCW mode. The minimum at around 84 K was highlighted by red circle and an arrow. (b) The inverse magnetic susceptibility data (FCW, 80–140 K, blue hollow circle) fitted with the modified Curie-Weiss model (orange line). (c) The Curie-Weiss fit was further extrapolated to 160 K.

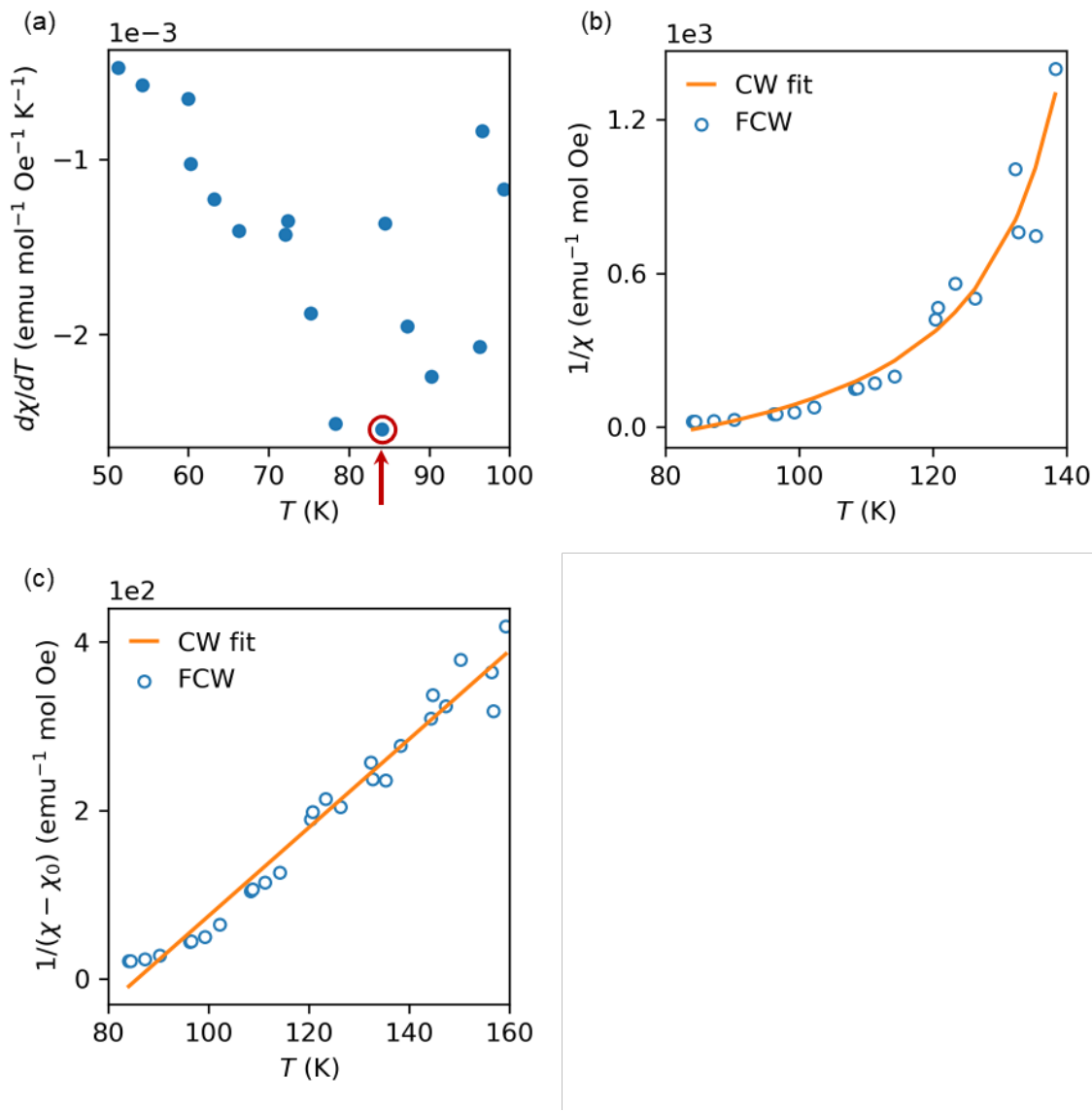

**Figure S3 Magnetic hysteresis.** Magnetic hysteresis observed in the high pressure $Sr_2IrO_4$ phase at 2 K ranging from -0.6 T to 0.6 T.

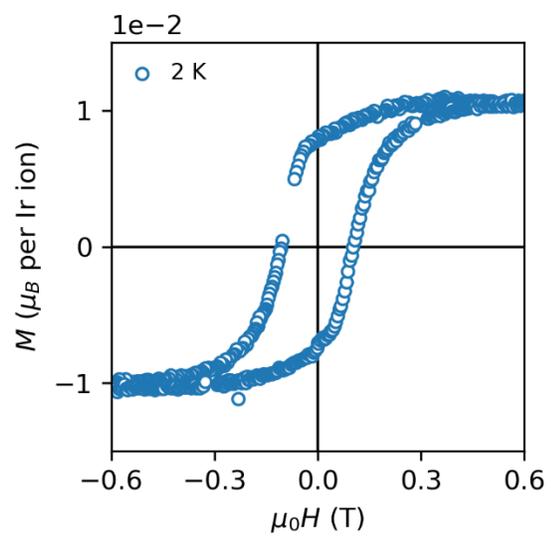

**Figure S4 Field dependence of specific heat.** Temperature dependence of specific heat data over temperature ($C_\mathrm{p}/T$) for high pressure Sr$_2$IrO$_4$ phase, under 0 T (blue), 0.05 T (orange), and 1 T (green). No significant differences were observed. No $\lambda$ shape anomalies emerged in the whole temperature regime studied under either case.

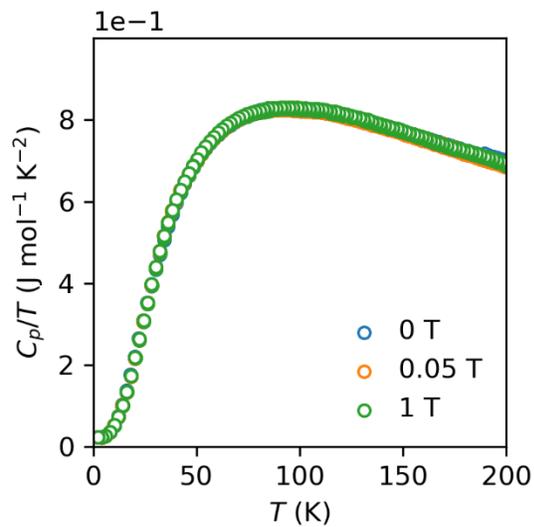

**Figure S5 Specific heat data fitted by Debye and Einstein model.** Temperature dependence of specific heat data over temperature under 0 T ($C_\mathrm{p}/T$, blue hollow circle) for high pressure $Sr_2IrO_4$ phase, fitted by (a) Debye model, (b) Einstein model, (c) two Debye model with the electronic contribution included, and weighted Debye model (d) without and (e) with the electronic contribution included.

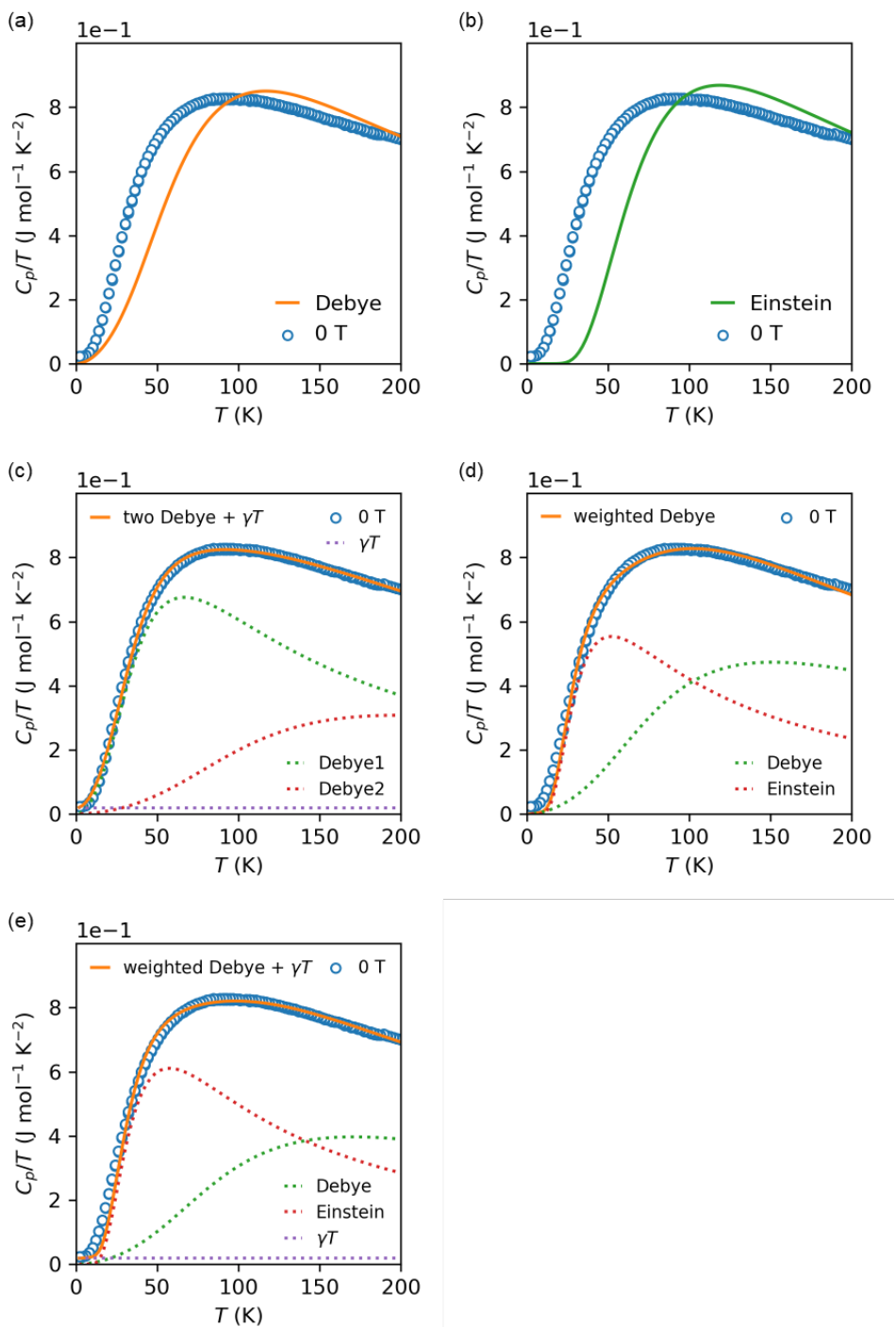

**Figure S6 Low temperature specific heat data (2–20 K).** Specific heat data over temperature ($C_\mathrm{p}/T$) plotted versus $T^2$ under low temperature regime, 2–20 K, providing the possibility to derivate the Sommerfeld parameter, $\gamma$.

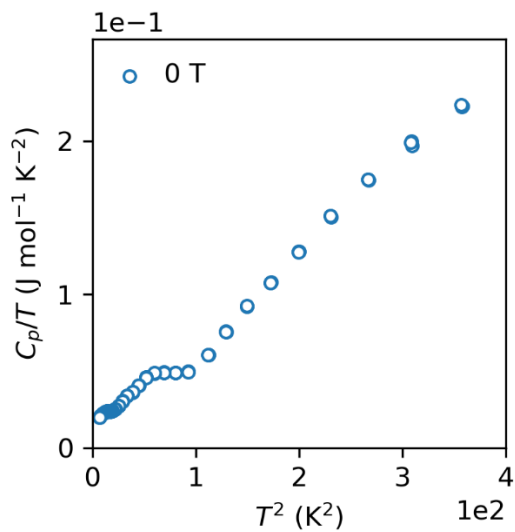

**Figure S7 Temperature dependence of resistivity.** Temperature dependence of the resistivity data $\rho$ plotted as $\ln \rho$ versus (a) $T^{-1}$, (b) $T^{-1/2}$, and (c) $T^{-1/4}$ under 0 T.

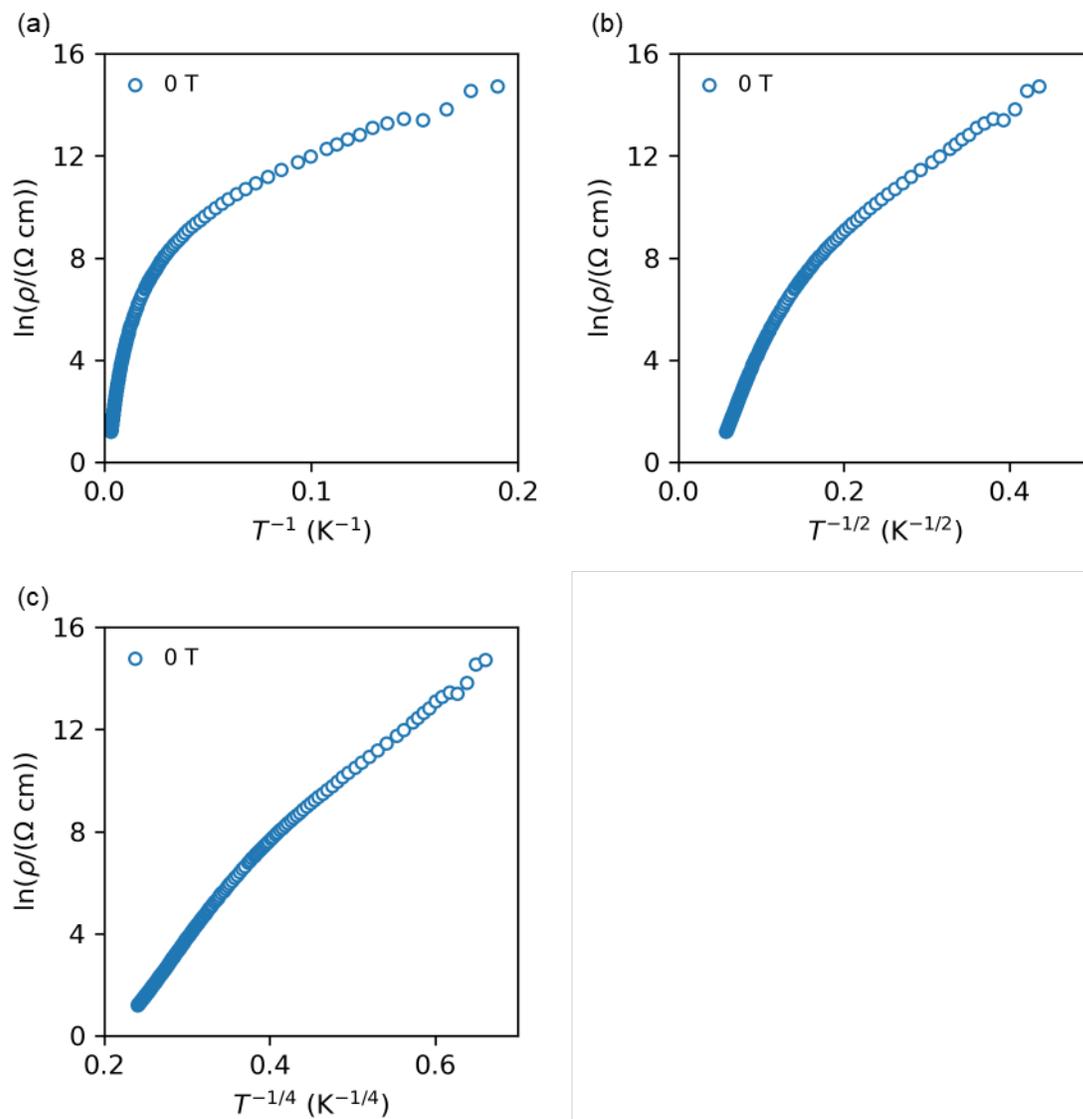

**Figure S8 Resistivity data fitted by Equation 9 with $v$ of 1/2 and 1/4.** (a) The resistivity data $\rho$ (blue hollow circle) ranging from 110–300 K fitted by **Equation 9** with $v$ of 1/2 (orange line). (b) The resistivity data $\rho$ in the low temperature regime ranging from 8–20 K fitted by **Equation 9** with $v$ of 1/2. (c) The resistivity data $\rho$ in the low temperature regime ranging from 10–20 K fitted by **Equation 9** with $v$ of 1/4. Fitting parameters were summarized in **Table S4**. The value $v$ of 1/4 is favored over 1/2.

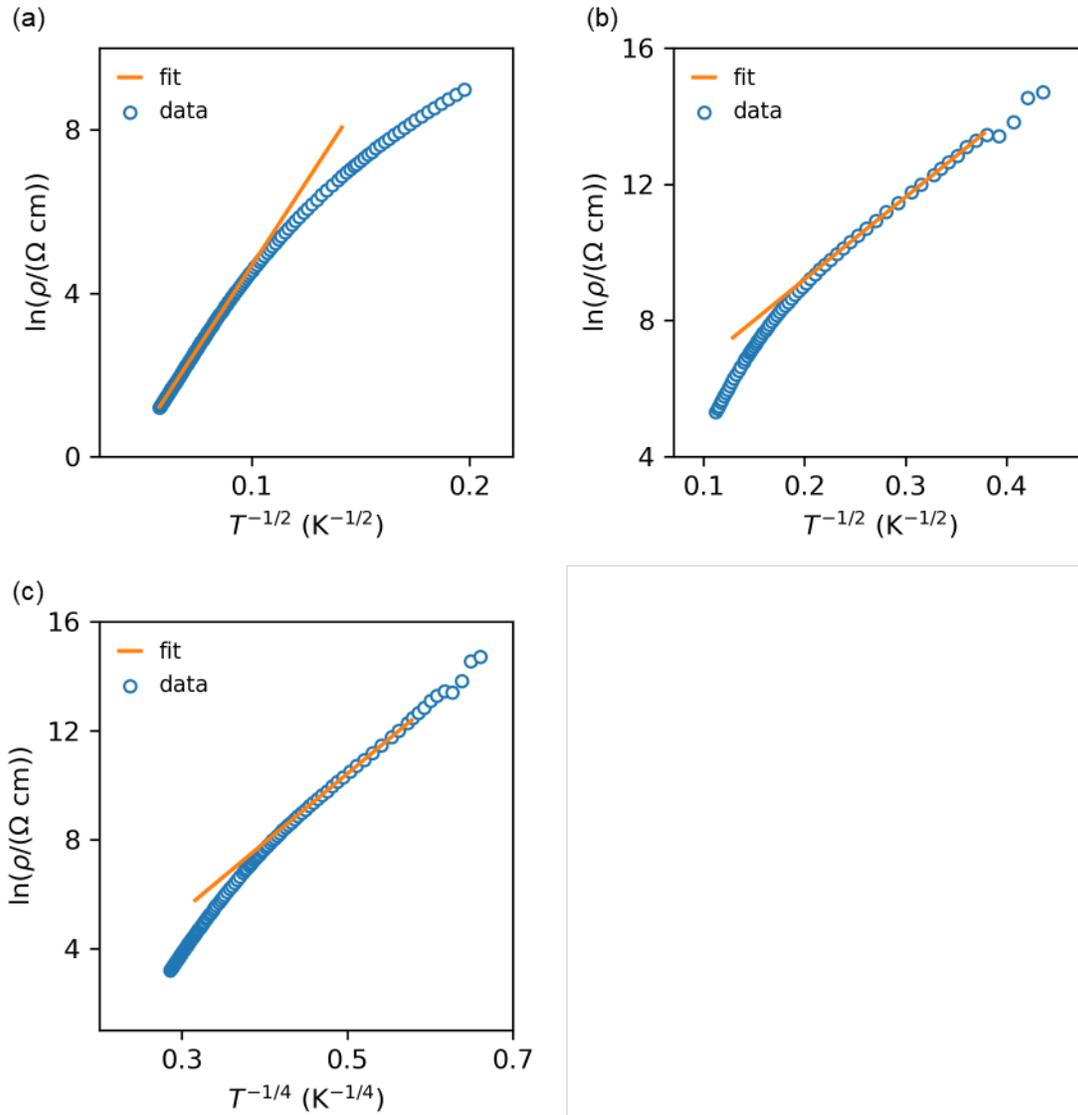

**Table S4. Summary of fitting parameters for resistivity data.** Summary of fitting parameters for the resistivity data $\rho$ by **Equation 9**. $R^2$ is the coefficient of determination.

| | $v = 1/2$ | | |
|---|---|---|---|
| Temperature Range / K | $\rho_0$ / ($\Omega$ cm) | $T_0$ / K | $R^2$ |
| 110–300 | 3.01(4) × 10$^{-5}$ | 6682 | 0.9996 |
| 8–20 | 79.6(37) | 583 | 0.9996 |

| | $v = 1/4$ | | |
|---|---|---|---|
| Temperature Range / K | $\rho_0$ / ($\Omega$ cm) | $T_0$ / K | $R^2$ |
| 80–300 | 7.82(12) × 10$^{-5}$ | 3.83 × 10$^6$ | 0.9998 |
| 10–20 | -2.23(4) | 4.10 × 10$^5$ | 0.9999 |